\documentclass[a4paper]{article}

\title{\textbf{Time-Variant Overlap-Add in Partitions}}

\author{Hagen Jaeger (1), Uwe Simmer (1), Jörg Bitzer (1), Matthias Blau (1) (2) \\ \\
	{\footnotesize (1)} {\small Jade Hochschule Oldenburg, Institut f\"ur H\"ortechnik und Audiologie} \\ {\small Ofener Stra{\ss}e 16, 26121 Oldenburg,  Germany} \\
	{\footnotesize (2)} {\small Cluster of Excellence "Hearing4All"} \\ {\small Ofener Stra{\ss}e 16, 26121 Oldenburg,  Germany} \\ \\
}

\date{\today}

\usepackage{todonotes}
\usepackage{url}
\usepackage{layout}
\usepackage[a4paper, left=3cm, right=3cm, top=3cm]{geometry}
\usepackage[skip=3pt plus1pt, indent=10pt]{parskip}
\usepackage{graphicx}
\usepackage{amsmath}
\usepackage{booktabs}

\newcommand{\tbl}[2]{
	\caption{#1}
	\relax
	#2
}
\let\oldrule\toprule
\renewcommand{\toprule}{\oldrule\oldrule}
\newcommand{\botrule}{\bottomrule}
\newcommand{\colrule}{\midrule}

\def \myfigsize {0.65\textwidth} 

\begin{document}
	
\maketitle

\begin{abstract}
	Virtual  and  augmented  realities are increasingly popular tools in
	many domains such as architecture, production, training and education,
	(psycho)therapy, gaming, and others. For a convincing rendering of
	sound in virtual and augmented environments, audio signals must be convolved in real-time with impulse responses that change from one moment in time to another. Key requirements for the implementation of such time-variant real-time convolution algorithms are short latencies, moderate computational cost and memory footprint, and no perceptible switching artifacts. In this engineering report, we introduce a partitioned convolution algorithm that is able to quickly switch between impulse responses without introducing perceptible artifacts, while maintaining a constant computational load and low memory usage. Implementations in several popular programming languages are freely available via GitHub.
\end{abstract}


\section{Introduction}

Fast convolution algorithms are used to digitally modify the spectral content of audio signals in real-time. Application examples include signal enhancement by noise reduction, or auralization for virtual acoustics. In these applications, it is important to ensure that (a) the extra delay introduced by the convolution procedure (input-output latency) is kept very small (typically less than 10~ms in virtual acoustics), (b) the impulse responses can be changed quickly without perceptible artifacts and (c) the computational load and memory footprint are kept within affordable limits.

The smallest input-output latency is obtained by using time-domain convolution.
However, time-domain convolution imposes a high computational load, to the extent that it often cannot be used with long and/or multichannel filter impulse responses.

In order to retain a small input-output latency while lowering the computational load, Allen and Rabiner~\cite{OLA} introduced the idea of replacing the time-domain based calculation by a block-oriented multiplication in the frequency domain, combined with overlapping addition of the convolution remains, known as Overlap-Add (OLA) procedure.
The OLA method was intended to perform time-invariant convolutions. If used with time-variant filters, it produces perceptible audio signal artifacts.

Crochiere et al.~\cite{WOLA} presented a procedure to overcome this issue by having the input blocks overlap by $50 \%$ and applying analysis and synthesis window functions when using the OLA scheme - this is known as Weighted Overlap-Add (WOLA).
WOLA suppresses the artifacts produced by time-variant filter functions, but the additional overlap increases the computational load.

Both OLA and WOLA perform block-oriented signal processing. 
Small block sizes lower the latency, but increase the computational load, and vice versa~\cite{Wefers}.
In most cases, 50\% overlap is used, in particular with WOLA.
For the sake of simplicity, we will stick to this value in the following considerations.

In order to further gain efficiency, Soo and Pang~\cite{POLA1, POLA2} introduced an OLA-based procedure which uses impulse response partitioning with equal or varying partition sizes.
This procedure is known as Partitioned Overlap-Add (POLA).
Still, POLA does not overcome the artifact issue when using time-varying transfer functions.
This can be achieved by using parallel POLA computations, the outputs of which are cross-faded (see e.g.~\cite{FWonder}).

In his work about real-time auralization, Wefers~\cite{Wefers} described and summarized a variety of calculation options for frequency-domain based fast convolution algorithms, and evaluated the efficiency of all available basic procedures such as OLA/POLA and the related Overlap-Save (OLS) and partitioned Overlap-Save (POLS) methods.
Based on these evaluations, the POLA procedure can be considered as a good choice for efficient real-time spectral modifications.
However, POLA- or POLS-based fast convolution methods with output cross fading can produce peaks in the computational load and have a considerable memory footprint.

In the following, we will present a procedure with constant computational load and optimized memory usage, which combines the efficiency of partitioned, block-oriented fre\-quency-domain convolution also supporting time-variant filter transfer functions by combining impulse response partitioning, frequency-domain Overlap-Add processing and time-domain signal block weighting in one procedure which we refer to as Time-Variant Overlap-Add in Partitions (TVOLAP).


\section{Method}
	
\begin{figure}[tbp]
	\centering
	\includegraphics[width=0.6\textwidth]{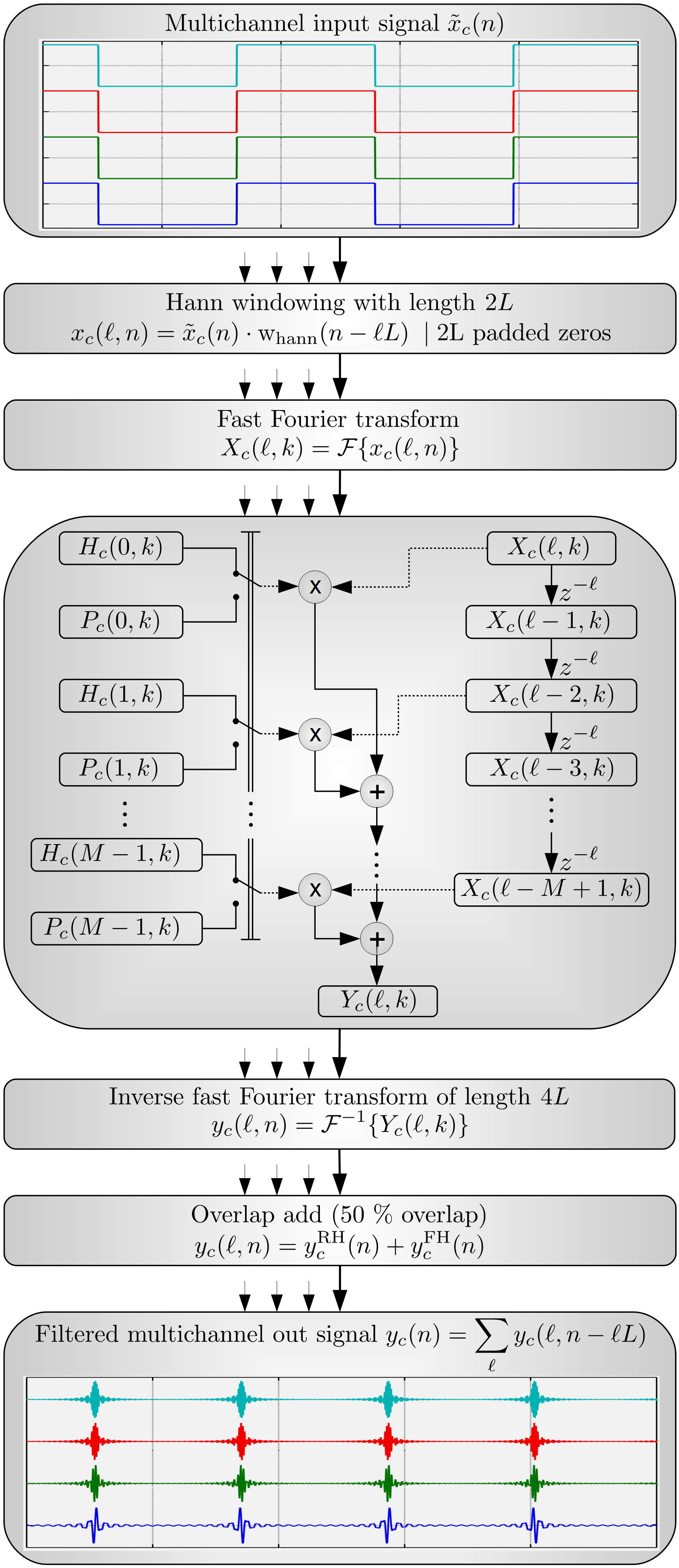}
	\caption{Schematic visualization of the TVOLAP signal processing stage, using a hop size of $L$ and a Hann window of length $2L$. In this example, transfer function partitions can be switched from $H(k,m)$ to $P(k,m)$ to illustrate how to deal with time-varying impulse responses.}
	\label{overallSchematics}
\end{figure}

TVOLAP is based on a partitioned overlap-add scheme.
Both the input signal and the impulse responses are divided into smaller blocks first:
Assuming a single-channel case, which can be extended to multiple channels easily, the input signal $\tilde{x}(n)$ is divided into blocks of length $2L$ every $L$ samples, with
\begin{equation}
	\hat{x}(\underbrace{n-\ell L}_{n'}, \ell) = \tilde{x}(n) \, \textrm{w}_{\textrm{hann}}(n-\ell L).
	\label{eq:1}
\end{equation}
Here, $\ell$ denotes the block index and $n$ the time index.
Each block is then weighted using a periodic Hann-window
\begin{equation}
	\textrm{w}_{\textrm{hann}}(n) = 1-\cos\left(2 \pi \frac{n}{2L}\right) \quad \textrm{for} \quad 0 \leq n < 2L ,
\end{equation}
resulting in windowed signal blocks that are overlapped by $50\%$, see also upper part of fig.~\ref{overallSchematics}

\begin{figure}[tbp]
	\centering
	\includegraphics[width=\myfigsize]{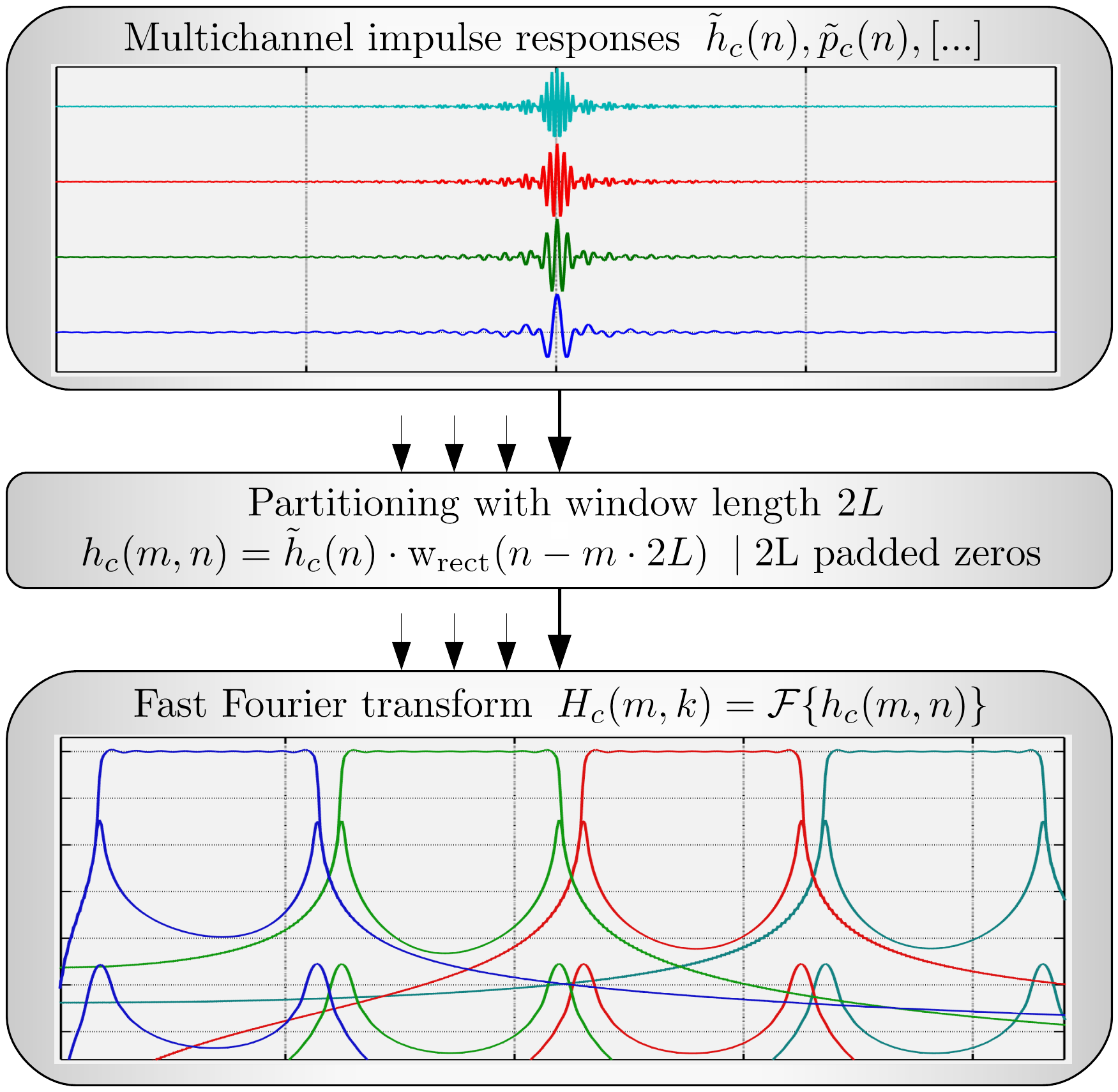}
	\caption{Example of computing transfer function partitions for the TVOLAP procedure. The four-channel impulse responses are partitioned into three non-overlapping partitions, each of length $2L$. Transfer function partitions are then obtained after zero-padding and Fourier transform.}
	\label{partSchematics}
\end{figure}

The impulse responses $\tilde{h}(n)$ are treated in a slightly different manner, see fig.~\ref{partSchematics}.
They are divided into $M$ {\em non-overlapping} partitions of length $2L$ with partition index $m$, i.e.
\begin{equation}
	\hat{h}(\underbrace{n-\ell \, 2L}_{n''} , m) = \tilde{h}(n) \, \textrm{w}_{\textrm{rect}}(n-m \, 2L),
	\label{eq:2}
\end{equation}
where  $\textrm{w}_{\textrm{rect}}$ is a length $2L$ rectangular window.   

Each input block $\hat{x}(n',\ell)$ and all impulse response partitions $\hat{h}(n'',m)$ are Fourier transformed after padding $2L$ zeros (double length zero padding), leading to a Fourier-transform length of $4L$.
In vector notation, we thus get input spectra
\begin{equation}
	X(k,\ell) = \sum_{n'=0}^{4L-1} x(n',\ell) \cdot e^{-j 2 \pi n' \frac{k}{4L}} ,
	\label{eq:5}
\end{equation}
with
\begin{equation}
	x(n',\ell) = [ \underbrace{\hat{x}(n',\ell)}_{2L}  \;\underbrace{ 0 \; 0 \; 0  \; ... \; 0}_{2L}],
	\label{eq:3}
\end{equation}
and transfer function partitions
\begin{equation}
	H(k,m) = \sum_{n''=0}^{4L-1} h(n'',m) \cdot e^{-j 2 \pi n'' \frac{k}{4L}} ,
	\label{eq:6}
\end{equation}
with
\begin{equation}
	h(n'',m) = [ \underbrace{\hat{h}(n'',m)}_{2L}  \; \underbrace{0 \; 0 \; 0  \; ... \; 0}_{2L} ] ,
	\label{eq:4}
\end{equation}
where $k$ denotes the frequency index in all equations.

In the frequency domain, intermediate output spectra $Y(k,\ell)$ for each block $\ell$ are computed by multiplying {\em every second} input spectrum with the transfer function partitions, 
\begin{equation}
	Y(k,\ell) = \sum_{m=0}^{M-1} H(k,m) \cdot X(k,\ell-2m).
	\label{eq:7}
\end{equation}
The corresponding time-domain signals are given by
\begin{equation}
	y(n',\ell) = \frac{1}{4L}\sum_{k=0}^{4L-1} Y(k,\ell) \cdot e^{j 2 \pi n' \frac{k}{4L}},
	\label{eq:8}
\end{equation}
each of length $4L$.

The final output signal is then calculated using a two-step overlap-add procedure.
In a first step, the right half of the preceding intermediate output signal block, denoted as
\begin{equation}
	y^{\textrm{PR}}(n',\ell) = y(n'+2L,\ell-1) 
	\label{eq:9}
\end{equation}
and the left half of the current intermediate output signal block, denoted as
\begin{equation}
	y^{\textrm{CL}}(n',\ell) = y(n',\ell) ,
	\label{eq:10}
\end{equation}
both with $0 \leq n' < 2L$, are summed, resulting in length $2L$ output signal blocks
\begin{equation}
	\hat{y}(n',\ell) = y^{\textrm{PR}}(n',\ell) + y^{\textrm{CL}}(n',\ell).
	\label{eq:11}
\end{equation}
In the second step, these output signal blocks are added in a time-aligned manner, i.e.\ shifted by $L$, to give the final output signal
\begin{equation}
	\tilde{y}(n) = \sum_{\ell} \hat{y}(\underbrace{n-\ell  L}_{n'},\ell).
	\label{eq:12}
\end{equation}
The whole procedure is summarized in fig.~\ref{overallSchematics}.


There are two noteworthy properties of this procedure:
First, time-variant impulse responses can easily be incorporated.
This is accomplished by simply exchanging the transfer function partitions $H(k,m)$
by new ones (denoted as $P(k,m)$ in fig.~\ref{overallSchematics}) in the spectral multiplication step (eq.~\ref{eq:7}).
It should be noted that {\em all} transfer function partitions must be exchanged at the same time.

Second, the procedure can easily be expanded to process $c$ channels in parallel.
This makes it suitable for processing multi-channel transfer functions in real time, such as HRTFs and BRIRs, or array-based individualized binaural renderings such as the ones proposed in~\cite{Rasumow2017}.


\section {Results}

In order to highlight the benefits of TVOLAP for time-variant convolution, we will first analyze the behavior of several algorithms at the moment of exchanging the impulse response and the artifacts resulting from this behavior.
Subsequently, the associated computational load will be calculated for all algorithms considered.

\begin{figure}[btp]
	\centering
	\includegraphics[width=\myfigsize]{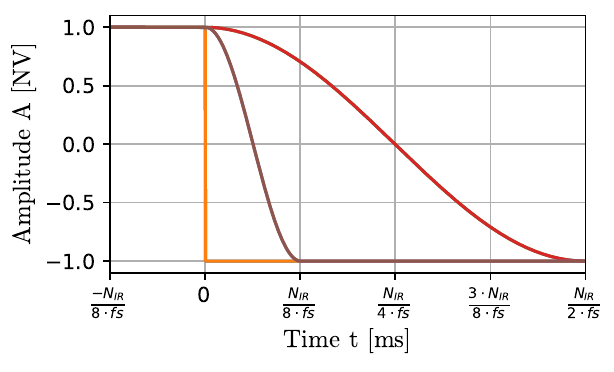}
	\caption{Switching behavior of OLA/OLS (rectangular), WOLA (wide cosine half) and TVOLAP (narrow cosine half), illustrated with a sequence of ones convolved with a delta impulse which switches its polarity at $t=0$. $N_\textrm{IR}$ denotes the impulse response length and the dependent block processing size. $M = 4$ partitions were used for TVOLAP.}
	\label{fig:rectSwitch}
\end{figure}

The switching behavior is analyzed by comparing the output that TVOLAP produces for specific input test signals and time-variant impulse responses to that of the well-known OLA, OLS and WOLA-procedures: 
In fig.~\ref{fig:rectSwitch}, the switching characteristics are shown for the case of a sequence of ones, convolved with a delta impulse the polarity of which is changed from plus one to minus one at $t=0$.

For OLA, as well as OLS, an abrupt change can be observed, while WOLA shows the typical Hann-window cross-fading between the two states. 

We use the switching latency to characterize the different procedures. It refers to the amount of time that a procedure needs to fully settle to the new state after the switching process is triggered.

The switching latency (assuming the sampling rate $f_\text{s}$) for WOLA equals $N_\textrm{IR}\big/ (2f_s)$ when a block size of $N_\textrm{IR}$ is used with 50\% overlap (the hop size then equals $N_\textrm{IR}/2$).
TVOLAP shows a similar switching characteristic, but with a  switching latency of $N_\textrm{IR} \big/ (2Mf_s)$. 

It should be noted that we assume a block size of $N_\textrm{IR}$ for OLA and OLS, which is the most commonly used choice, although other choices are possible. Still, implementations with smaller block sizes are less efficient, because the amount of required fast Fourier transforms increases, while their length does not decrease.
Usually, impulse response partitioning is therefore employed to achieve smaller latencies.

\begin{figure}[btp]
	\centering
	\includegraphics[width=\myfigsize]{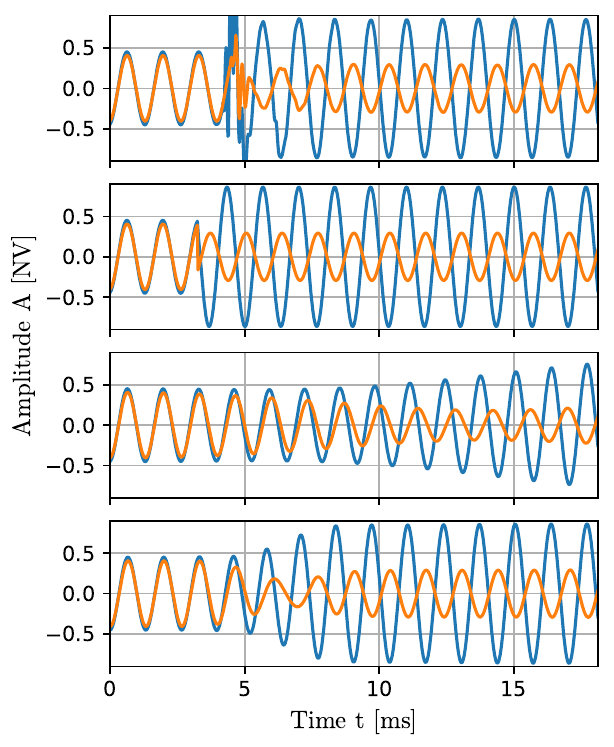}
	\caption{Output of OLA, OLS, WOLA and TVOLAP (from top to bottom) when a 750~Hz sine is convolved with the TU Berlin Kemar HRIRs, switched from 
		$0^\circ$ to $90^\circ$ azimuth ($0^\circ$ elevation) at $t=3 \, \textrm{ms}$.
		Light gray - left ear signal, dark grey - right ear signal.}
	\label{fig:sineSwitch}
\end{figure}

In fig.~\ref{fig:sineSwitch}, the switching characteristics of OLA, OLS, WOLA and TVOLAP (from top to bottom) are shown for the case of a $750\,$Hz sine convolved with the head-related impulse responses (HRIRs) of the KEMAR artificial head, measured by TU Berlin~\cite{TUB}, at $0^\circ$ azimuth (frontal incidence) and an abrupt switch to $90^\circ$ azimuth (incidence from right) at $t=3 \, \textrm{ms}$.
The block size for OLA, OLS and WOLA was set to $N_\textrm{IR}=2048$ (length of the HRIRs).
For TVOLAP, a block size of $2L=512$ and $M=4$ partitions were used, respectively.
The frequency of $750\,$Hz was chosen because the acoustic distance between the two KEMAR ears is close to half the wavelength at $750\,$Hz.
Hence at $90^\circ$ sound incidence one would expect a $180^\circ$ phase shift and a clear level difference between the left and right ear signals, whereas they should be very similar at frontal incidence.
This behavior of the initial and the final states can be seen with all algorithms.
They differ, however, in the way the transition between initial and final state is accomplished:

OLA shows a high amount of distortions after a small, channel-dependent delay following the switching event.
This delay is caused by the characteristics of the processing routine which adds the remainder of the last block, calculated with the old impulse response, to the current output, calculated with the new impulse response.
OLS does not use remaining outputs, and therefore does not show any distortions. Due to its property to use only the spectrum parts which are free of circular convolution artifacts, it immediately delivers the new output, resulting in a hard switch between the two states.
This however produces audible artifacts, perceived as `clicking' sound .
In contrast, WOLA shows a soft crossover between the two states, resulting in a continuous and smooth switching procedure. TVOLAP shows the same behavior as WOLA, but uses only one fourth of the switching time and therefore generates the output signal with the new states faster.

\begin{figure}[btp]
	\centering
	\includegraphics[width=\myfigsize]{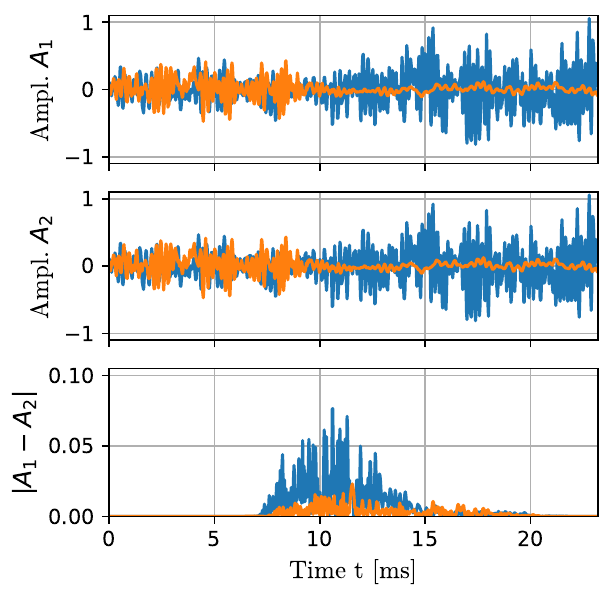}
	\caption{Comparison of the output of TVOLAP to a time-domain convolution with output cross-fading (CF-TDC) procedure. A pink noise was convolved with the TU Berlin KEMAR HRIRs, which were switched from $0^\circ$ to $90^\circ$ azimuth ($0^\circ$ elevation) at $t=7 \, \textrm{ms}$. Top: TVOLAP, middle: CF-TDC, bottom: absolute difference between TVOLAP and CF-TDC.
		Light gray - left ear signal, dark grey - right ear signal.}
	\label{fig:errEval}
\end{figure}

Time-domain cross-fading of two output streams is a well-known practice to suppress perceptible switching artifacts.
TVOLAP was compared against two independently calculated and subsequently cross-faded time domain convolutions (CF-TDC). Fig.~\ref{fig:errEval} shows the output signals when pink noise is convolved with the same $0^\circ$ and $90^\circ$ KEMAR HRIRs as above, using either TVOLAP (top diagram) or CF-TDC (middle diagram).
Differences between the two approaches are present but small - they become only visible when the difference signals are analyzed (bottom diagram).

Thus, TVOLAP results in a similarly smooth transition behavior as CF-TDC.
However, CF-TDC is a highly inefficient method since it doubles the necessary computations of the already inefficient time-domain convolution method in order to compute the second stream continuously. If the second audio stream was computed around the switching moment only, the computational cost would be heavily fluctuating over time with peaks during switching.

In order to theoretically assess the computational efficiency, the number of floating point operations per second (FLOPS) needed for the block-oriented processing is compared for four algorithms, namely: the direct convolution with equally long output signals, OLA, WOLA and TVOLAP. The exemplary calculation first assumes an impulse response $\tilde{h}(n)$ of length $N_{\textrm{IR}} = 2048$ and an audio sample rate of $f_s = 48\, \textrm{kHz}$, see table~\ref{tab:01} for aggregated results.

In the direct convolution approach, the output signal $\tilde{y}(n)$ is computed as
\begin{equation}
	\tilde{y}(n) = \sum_{i=0}^{N - 1} \tilde{x}(i) \cdot \tilde{h}(n-i)
	\label{eq:13}
\end{equation}
with $i$ being the shifting index. The procedure needs $N \cdot f_s$ multiplications and additions to calculate one second of the output signal, resulting in $2 \cdot 2048 \cdot 48000 = 196608000 \,$ $\textrm{FLOPS} = 196.608 \, \textrm{MFLOPS}$ in case of $N=2048$.
It is possible to perform a soft crossover between two switching states with the above-mentioned CF-TDC method, which would need additional calculations, as the number of FLOPS needed will double around the switching time, plus two additional multiplications and one addition per audio sample, resulting in $3 \cdot f_s = 0.144$ extra MFLOPS.
Furthermore, a switching latency which equals the cross-fading time is introduced.
The objective of this procedure is to not introduce any latency onto the audio stream, while the switching latency can be selected without any restrictions.
Since this convolution approach directly works in the time domain, it is extremely inefficient.

For the exemplary OLA routine, a block size of $N_{\textrm{IR}}$ ($ = 2048$) is chosen, as it is the most widely used implementation choice. A real-valued fast Fourier transform of length $2 \cdot N_\textrm{IR}$ (zero padding up to twice the block size), implemented with the decimation-in-time scheme, needs $N_\textrm{IR}\big/2 \cdot \textrm{log}_{2}(N_\textrm{IR})$ butterflies, plus an overhead of $4 \cdot N_\textrm{IR}$ multiplications and $10 \cdot N_\textrm{IR}$ additions.
A single butterfly operation consists of 4 real-valued multiplications and 6 real-valued additions, resulting in an overall amount of $10 \cdot N_\textrm{IR}\big/ 2 \cdot \textrm{log}_{2}(N_\textrm{IR}) + 14 \cdot N_\textrm{IR}$ arithmetic operations per fast Fourier transform, or $5 \cdot \textrm{log2}(N_\textrm{IR}) + 14$ arithmetic operations per sample (calculation also applies to the inverse transform).
The OLA procedure consists of the fast Fourier transform plus its inverse, $N_\textrm{IR}(+1)$ complex-valued multiplications (6 arithmetic operations each) and $N_\textrm{IR}$ additions for each block, or $2 \cdot (5 \cdot \textrm{log2}(2048) + 14) + 6 + 1 = 145$ arithmetic operations per audio sample, resulting in $145 \cdot 48000 = 6.96 \, \textrm{MFLOPS}$.
It can easily be seen that OLA is much more efficient than the direct convolution approach (approximately by a factor of 28 for this example), but it introduces a latency of $N_\textrm{IR}$ samples to the audio stream.
Also, it does not support time-variant filtering by default.

The amount of FLOPS needed for the exemplary WOLA routine, assuming again a block size of $N_{\textrm{IR}} = 2048$ and square root Hann analysis and synthesis windows, can be calculated in much the same way. 
The only difference to OLA is that for WOLA, each audio sample has to be transformed, multiplied and added two times, which is due to the default 50\% input signal block overlap. Additionally, two extra multiplications and one addition per sample are needed to apply the analysis and synthesis windows, as well as the overlapping output block summation.
All in all, this results in $2 \cdot (145 + 3) \cdot 48000 = 14.064 \, \textrm{MFLOPS}$ for WOLA, approximately twice the computational load compared to OLA.
WOLA also introduces a latency of $N_\textrm{IR}$ samples to the audio stream, but it is suitable to process time-variant filters.
The switching latency equals half the block size and hence depends on the impulse response length.

The exemplary TVOLAP routine works with a block size of $2L=512$ and therefore uses $M = N_{\textrm{IR}}\big/ (2L) = 4$ impulse response partitions.
Similarly to WOLA, each audio sample has to be transformed back and forth, multiplied and added two times, due to the 50\% input block overlap.
In contrast to WOLA, only one Hann analysis window is used.
Summing up all processing states, this results in $2 \cdot 2 \cdot (5 \cdot \textrm{log2}(512) + 14)$ arithmetic operations for the back and forth transforms, plus $2 \cdot M = 8$ complex-valued multiplications and additions ($\widehat{=} \, 8 \cdot 8$ arithmetic operations) as well as $2 \cdot 2 \cdot 1$ multiplications for windowing the input and for the overlapping output block summation, per audio sample, respectively.
This yields an amount of $(20 \cdot \textrm{log2}(512) + 56 + 64 + 4) \cdot 48000 = 14.592 \, \textrm{MFLOPS}$ for the TVOLAP procedure,
which is similar to WOLA for this example.
However, only a quarter of the audio and switching latencies are needed as a consequence of the partitioning process.
The exemplary TVOLAP routine is still approximately 13 to 14 times more efficient than the CF-TDC approach.
In addition, TVOLAP does not introduce peaks in the computational load when changing the impulse response, as only the transfer function partitions are exchanged (see fig.~\ref{overallSchematics}).

In a second example, a shorter impulse response of length $N_{\textrm{IR}} = 512$ is considered, see table~\ref{tab:02}. OLA is approximately eight times faster than the direct convolution approach, which is much less effort in computational load compared to the processing with $N_{\textrm{IR}} = 2048$ (Table \ref{tab:01}). This emphasizes the fact that OLA gains most in efficiency when using high block processing sizes, which on the other hand increases the audio latency. WOLA is approximately half as efficient as OLA for the already explained reasons (input block overlap). In addition, it can be seen that WOLA and TVOLAP require the same amount of MFLOPS, with the exception of the additional synthesis window application for WOLA, if $M=1$. This is because TVOLAP can be described as WOLA with a Hann analysis and a rectangular synthesis window if only one partition is used.
\begin{table}
	\tabcolsep8.1pt
	\centering
	\tbl{Computational cost of CF-TDC, OLA, WOLA and TVOLAP, measured in FLOPS, when processing an impulse response of length $N_{\textrm{IR}} = 2048$. The `$>$' symbol denotes the extra computational load and switching latency for CF-TDC, which depend on the chosen cross-fading time. The block size for TVOLAP was set to 512. \vspace{0.15cm}}{%
		\begin{tabular}{@{}lccc@{}}\toprule
			Algorithm & audio lat. & switching lat. & MFLOPS \\
			$N_\textrm{IR} = 2048$ & & & \\\colrule
			CF-TDC & 0 & $>$0 & $>$196.608 \\
			OLA & 2048 & 0 & 6.96 \\
			WOLA & 2048 & 1024 & 14.064 \\
			TVOLAP & 512 & 256 & 14.592 \\\botrule
	\end{tabular}}
	\label{tab:01}
\end{table}
\begin{table}
	\tabcolsep8.1pt
	\centering
	\tbl{Computational cost of CF-TDC, OLA, WOLA and TVOLAP, measured in FLOPS, when processing an impulse response of length $N_{\textrm{IR}} = 512$. The `$>$' symbol denotes the crossfading time dependent extra computational load and switching latency for CF-TDC, which depend on the chosen cross-fading time. The block size for TVOLAP was set to 512. \vspace{0.15cm}}{%
		\begin{tabular}{@{}lccc@{}}\toprule
			Algorithm & audio lat. & switching lat. & MFLOPS \\
			$N_\textrm{IR} = 512$ & & & \\\colrule
			CF-TDC & 0 & $>$0 & $>$49.152 \\
			OLA & 512 & 0 & 6.0 \\
			WOLA & 512 & 256 & 12.288 \\
			TVOLAP & 512 & 256 & 12.192 \\\botrule
	\end{tabular}}
	\label{tab:02}
\end{table}

In summary, out of the methods that are capable of handling time-variant filters without perceptible artifacts (CF-TDC, WOLA, and TVOLAP), CF-TDC exhibits the lowest latencies at the cost of high computational loads, in particular when impulse response lengths are getting longer.
In the case of WOLA, the block size controls the latencies.
Since it is determined by the impulse response length, WOLA will show high latency values and a slow cross fading behavior for long impulse responses like BRIRs.
TVOLAP in contrast separates the dependency between block size $L$ and the impulse response length by partitioning, which enables it to convolve long, time-variant impulse responses with low latency, short switching delays and good overall-efficiency.
	

\section{Application Example}

\begin{figure}[h!btp]
	\centering
	\includegraphics[width=\myfigsize]{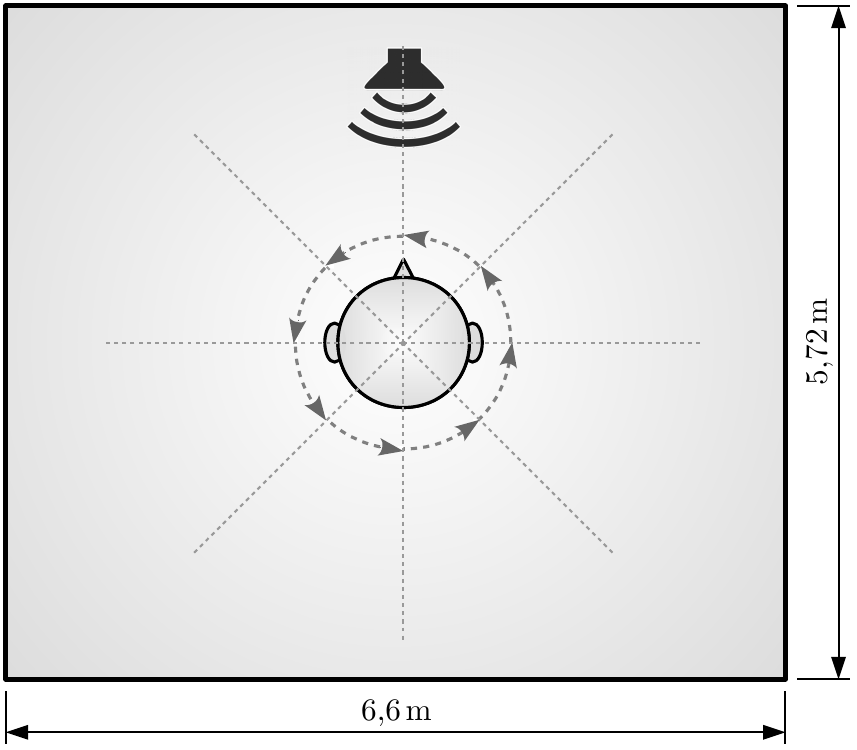}
	\caption{Visualization of a BRIR switching experiment. The acoustic source (illustrated as black loudspeaker symbol) was statically centered in front of the listener, while the listener rotated counterclockwise in steps of 45$^\circ$. The switching was intended to happen immediately. Positions (x,y,z): Listener centered at ($3.3 \,$m, $2.86 \,$m, $1.06 \,$m), Source centered at ($3.3 \,$m, $4.96 \,$m, $1.06 \,$m), `Shoe-box' room geometry with ($6.6 \,$m, $5.72 \,$m, $2.8 \,$m).}
	\label{fig:experiment}
\end{figure}

As an application example, we want to look at the case of processing time-variant binaural room impulse responses (BRIRs).
Fig.~\ref{fig:experiment} describes the experiment and its overall settings.
In a virtual scene, a room (reverberation time approx. 0.65~s) was rendered with a loudspeaker in front of the listener.
The listener rotated counterclockwise in steps of 45 degrees while a signal was played via the loudspeaker.
The rendering was realized by using BRIR set `x0y0'
from the Salford-BBC Spatially-sampled Binaural Room Impulse Responses (SBSBRIR)~\cite{SBSBRIR}.
The BRIRs have a length of $N_{\textrm{IR}} = 32.768$ samples at a sampling rate of $f_s = 44,1\,$kHz.
Hence, the block size for OLA, OLS and WOLA was set to $L_{1} = N_{\textrm{IR}} = 32.768$, whereas TVOLAP used a block size of $L_{2} = 1024$ and $M = 32$ partitions, respectively.
The BRIRs were switched every $N_{\textrm{IR}}$ samples for OLA, OLS and WOLA, in a way that every new block was convolved with a new BRIR.
With TVOLAP, the BRIRs were switched after $32$ blocks, in order to get the same switching start time as with the other algorithms.
The resulting output can be listened to at: \\

\url{https://tgm-oldenburg.github.io/TVOLAP/}. \\

The anticipated artifacts are clearly audible; OLA and OLS produce perceptible click sounds, while WOLA performs a smooth cross-fade of the states.
This does not introduce distortions, but takes too much time to be able to follow the intended discrete switching of the spatial direction.
TVOLAP is closest to the intended scene, the sound changes its perceived position immediately to the next discrete loudspeaker position, free of perceptible artifacts.


\section{Conclusions}

In this paper, we introduced an extension of the class of partitioned fast convolution algorithms by especially addressing time-variant scenarios. However, it is difficult to reproduce the natural behavior for the processing of time-variant scenarios, e.g. auralizing the real behavior for a switching between several processing states. Therefore, the selection of an optimal algorithm is typically application-driven.
The proposed time-variant overlap-add in partitions (TVOLAP) processing routine delivers a good trade-off between high computational efficiency and short audio and switching latencies without introducing noticeable switching artifacts.
The latencies can be adjusted within reasonable limits, without excessively compromising computational efficiency.
It therefore provides a flexible framework for fast convolution tasks.

To the authors' knowledge, no universally accepted procedures to perceptually determine the ground truth in time-variant scenarios, such as auralization of dynamically changing acoustic environments, exist - which may be a topic of further research. Because of its flexibility, the proposed method might be handy in such work.

To simplify its use, TVOLAP was implemented in several common computer languages, namely Python, C++ and Matlab/Octave, and is freely available at GitHub~\cite{TVOLAP} under the restrictions of a lesser general public license (LGPL). 



\section{Acknowledgement}

Parts of this work were supported by Bundesministerium für
Bildung und Forschung under grant no.~03FH021IX5 and by Deutsche Forschungsgemeinschaft
(DFG, German research Foundation), project ID 444832396,
within SPP 2236 Audictive.


\newpage


\centering{{\Large \textbf{The Authors}}}

\vspace{0.025\linewidth}

\begin{figure}[h!]
	\centering
	\begin{minipage}[b]{0.2\linewidth}
		\includegraphics[width=\linewidth,clip,keepaspectratio]{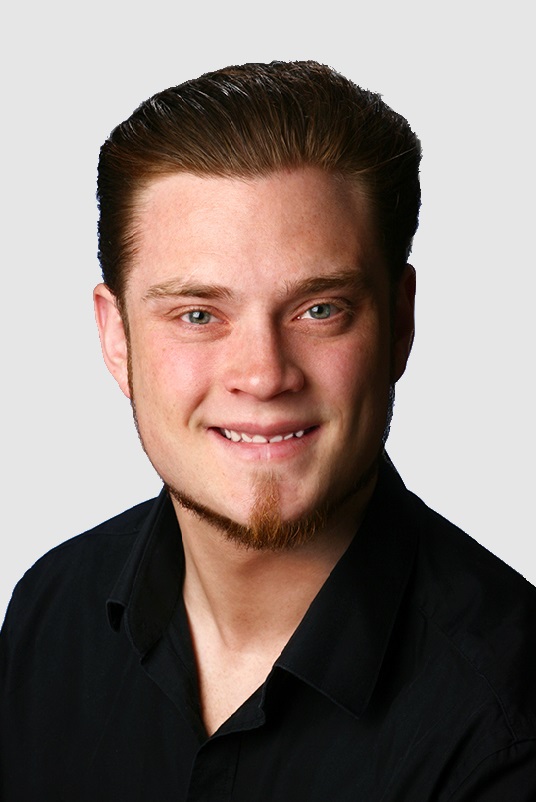}
		\centering{\large Hagen Jaeger}
	\end{minipage}
	\hspace{0.05\linewidth}
	\begin{minipage}[b]{0.2\linewidth}
		\includegraphics[width=\linewidth,clip,keepaspectratio]{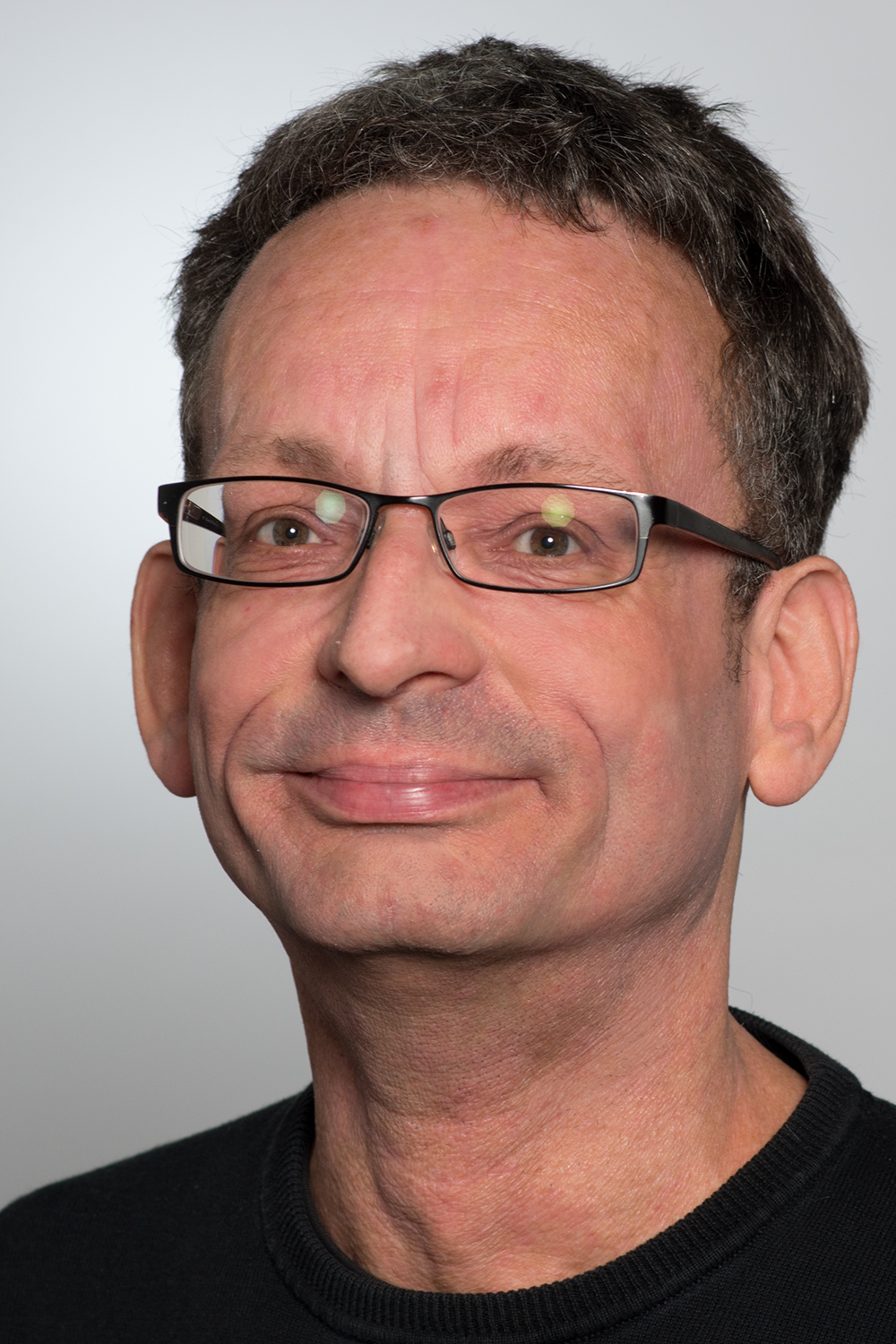}
		\centering{\large Uwe Simmer}
	\end{minipage}
	\hspace{0.05\linewidth}
	\begin{minipage}[b]{0.2\linewidth}
		\includegraphics[width=\linewidth,clip,keepaspectratio]{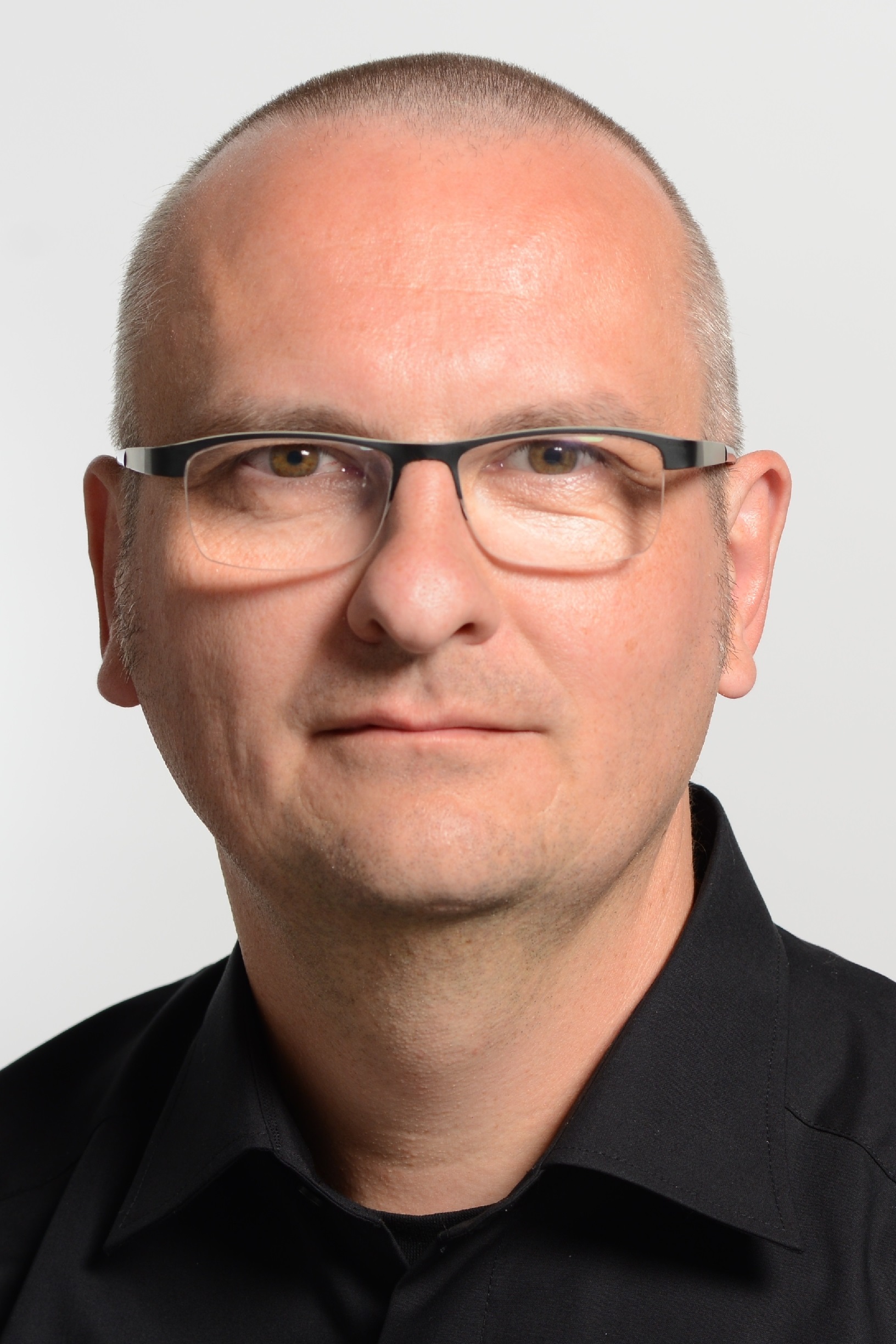}
		\centering{\large Jörg Bitzer}
	\end{minipage}
	\hspace{0.05\linewidth}
	\begin{minipage}[b]{0.2\linewidth}
		\includegraphics[width=\linewidth,clip,keepaspectratio]{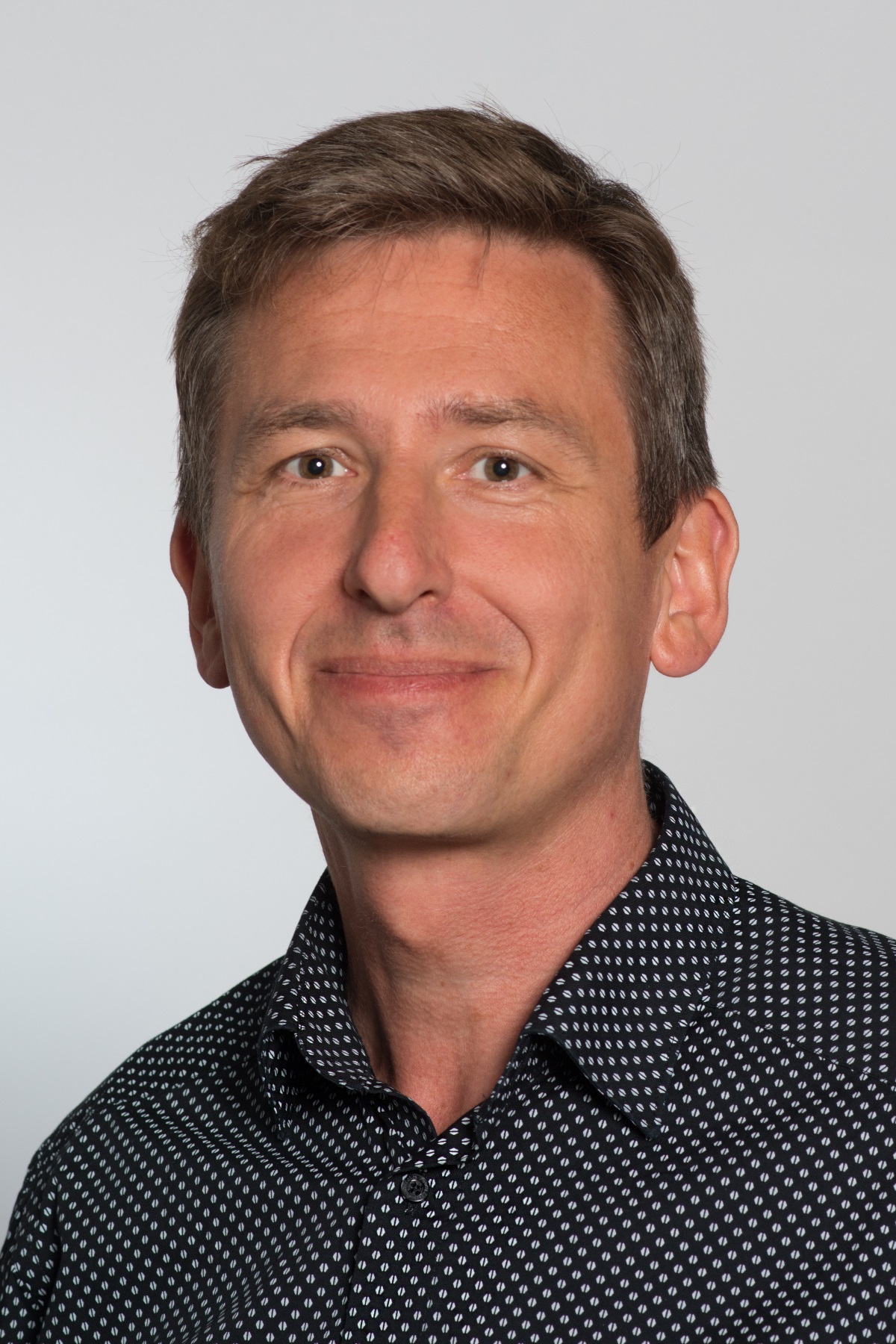}
		\centering{\large Matthias Blau}
	\end{minipage}
\end{figure}

\vspace{0.025\linewidth}

\begin{minipage}{\textwidth}
	\textbf{Hagen Jaeger} studied hearing technologies and audiology at the Jade University of Applied Sciences and University of Oldenburg (Germany), where he received his M.Sc. degree in 2017. Until that time, he worked in parallel as part of the Fraunhofer IDMT Hearing4All research cluster on the development of fixed point signal processing algorithms for embedded systems, which also was a subject of his bachelor thesis in 2015. From 2017 to 2021, he was part of the Fraunhofer IAIS department NetMedia, which addresses several engineering and research topics on the field of signal processing, augmented reality and big data analytics. Currently, he is working in the automotive industry for CARIAD SE.
\end{minipage}

\vspace{0.2in}

\begin{minipage}{\textwidth}
	\textbf{Uwe Simmer} received his Diploma and Ph.D. degrees both from the University of Bremen Germany in 1986 und 1988, respectively. In 1994 he joined Houpert Digital Audio being responsible for the development of algorithms for a binaural digital hearing aid. In 1997 he joined Aureca Audio-Systemtechnik GmbH. From 2002 to 2012 he was the founder, CEO and staff of ASP Acoustic Signal Processing GmbH, a company developing noise reduction and echo cancellation software for a car supplier. Since 2005 he is a member of the scientific staff of the Jade University of Applied Science Wilhelmshaven / Oldenburg / Elsfleth. Dr. Simmer is a member of the Acoustical Society of America and the Audio Engineering Society.
\end{minipage}

\vspace{0.2in}

\begin{minipage}{\textwidth}
	\textbf{Jörg Bitzer} received his diploma in 1995 and his doctorate in electrical engineering in 2002 from the University of Bremen where he also worked as a research assistant until 1999. From 2000 to 2003 he was head of the algorithm development team at Houpert Digital Audio, a  company specialized in  audio signal processing. Since September 2003 he is a professor for audio signal processing  at the Jade  University of Applied Science Wilhelmshaven/Oldenburg/Elsfleth. In 2010 he joined the Fraunhofer project group for hearing, speech, and audio technology in Oldenburg as a scientific supervisor. His current research interests include all forms of single- and  multichannel speech enhancement, audio restoration, audio effects for musical applications, and information retrieval for large media archives.
\end{minipage}

\vspace{0.2in}

\begin{minipage}{\textwidth}
	\textbf{Matthias Blau} studied electrical engineering at TU Dresden (Germany) and Purdue University (USA). In 1999 he received a Ph.D. in electrical engineering from TU Dresden. Since 2003 he is Professor of Electroacoustics at Jade Hochschule in Oldenburg, Germany. His current research interests include the acoustics of outer and middle ears, virtual acoustics, implantable transducers, room acoustics, and acoustical measurements. Dr. Blau is member of the German Acoustical Society, the Acoustical Society of America, and the Marie Curie Fellowship Association.
\end{minipage}
	
\end{document}